# Brain-Computer Interface Technologies in the Coming Decades

Brent J. Lance[+], Scott E. Kerick, Anthony J. Ries, Kelvin S. Oie, Kaleb McDowell[+]*

Translational Neuroscience Branch, Army Research Laboratory, Aberdeen Proving Ground, Maryland, USA.

[+] Brent J. Lance and Kaleb McDowell contributed equally to this work.

* Corresponding author:

Kaleb McDowell, IEEE Senior Member

Chief, Translational Neuroscience Branch

Army Research Laboratory

Aberdeen Proving Ground, MD, USA

Email: kgm8@cornell.edu

**Abstract:** As the proliferation of technology dramatically infiltrates all aspects of modern life, in many ways the world is becoming so dynamic and complex that technological capabilities are overwhelming human capabilities to optimally interact with and leverage those technologies. Fortunately, these technological advancements have also driven an explosion of neuroscience research over the past several decades, presenting engineers with a remarkable opportunity to design and develop flexible and adaptive brain-based neurotechnologies that integrate with and capitalize on human capabilities and limitations to improve human-system interactions. Major forerunners of this conception are brain-computer interfaces (BCIs), which to this point have been largely focused on improving the quality of life for particular clinical populations and include, for example, applications for advanced communications with paralyzed or "locked-in" patients as well as the direct control of prostheses and wheelchairs. Near-term applications are envisioned that are primarily task-oriented and are targeted to avoid the most difficult obstacles to development. In the farther term, a holistic approach to BCIs will enable a broad range of task-oriented and opportunistic applications by leveraging pervasive technologies and advanced analytical approaches to sense and merge critical brain, behavioral, task, and environmental information. Communications and other applications that are envisioned to be broadly impacted by BCIs are highlighted; however, these represent just a small sample of the potential of these technologies.

1. **Introduction**

Envision technologies that increase training or rehabilitation effectiveness by integrating real-time brain activity assessment into individualized, adaptive training and rehabilitation regimens; technologies that help you focus or even overcome a bad day by adjusting your environment to help you achieve desired brain states; technologies that help your doctor identify brain-based diseases or disorders before they interfere with life by assessing neural activity before behavioral symptoms appear; or even technologies that help you communicate better by assessing the neural activity of your audience and providing suggestions for increased clarity and interest. These are examples of potential brain-computer interface (BCI) technologies, a class of neurotechnologies originally developed for medical assistive applications. While there are a number of potential definitions for this term, in this paper we will expand the term BCI to include all technologies that use on-line brain-signal processing to influence human interactions with computers, their environment, and even other humans. This field that has recently seen an explosion of research enabled by recent advances in wearable, mobile biosensors and data acquisition; neuroscience; computational and analytical approaches; and computing for mobile brain imaging, all of which are enabling potential BCI applications that expand well beyond those initially developed for clinical populations. Further, these technologies, when combined with advancements in other fields such as pervasive computing, will push applications beyond human-computer interfaces and into the very nature of how people interact with computers and their environment. Over the next decades, brain-based technologies will allow computers, for the first time, to leverage sophisticated analyses of the emotional and cognitive states and processes of the people using them, revolutionizing the basic interactions people have, not only with the systems they use, but also with each other.

## 2. Background

Computers touch almost every aspect of our lives, performing critical functions in diverse areas including education and training, home and entertainment, medicine, and work. The importance of computers in our lives makes human-computer interaction one of the most critical factors in systems design. One fundamental issue in human-computer interaction is that limitations exist on the communication between human and computer. That is, human-system interaction is still fundamentally bounded by the inherent capabilities of humans to absorb, analyze, store, and interpret information to create behavior; and by limitations in the ability of computers to predict human intentions, action, and communications. Over the past decades, tremendous advancements have pushed the bounds on these limitations, including the development of novel devices for improving information flow into the computer via multi-modal devices [1], [2], collaborative performance among groups of people [3], eye trackers [4], speech and language [5], touch screens, gesture, and motion capture [6], [7], and facial expression recognition [8]; and for allowing the computer to provide more useful, relevant, or realistic information back to the user through improved visual displays and graphics, tactile and haptic feedback [9], [10], 3D audio [11], and virtual reality (VR) environments [12]. In addition, improved algorithmic approaches for predicting human behavior and intention, such as collaborative filtering [13], physiological computing [14], affective computing [15], user modeling [16], [17], and player modeling [18], open up the possibility of adapting devices to users and their needs. These steps have increased the quantity, quality, and interpretation of information transferred between the human and the system; however, as computational capabilities and complexity increase, the limited bandwidth between human and machine will become increasingly constraining. The tremendous growth of research in the field of neuroscience over the past several decades offers an approach to address these limitations. This basic research offers many potential insights into the brain state and mental processes of the human; insights that could potentially expand the current fundamental bounds on human-computer communications and open the door to completely novel approaches to both human-computer and human-human interaction [19], [20].

Major forerunners of these future brain-based technologies are early BCIs, which were intended to provide a direct communication pathway between the human brain and an external device. First developed in the 1970s [21], but largely unexplored until the past decades, early BCIs predominately focused on improving the quality of life of particular clinical populations. Exemplars of these early BCIs include devices for providing advanced communications with paralyzed or "locked-in" patients and the direct control of prostheses and wheelchairs.

*Early Approaches to BCI*

Two methodological constraints defined the nature and the scope of early BCI applications. The first constraint was to require users to focus on a particular task. For example, a typical application for spelling and writing had users focus on a single letter while watching streams of letters presented by the computer. These letters induce event-related potentials (ERPs) such as P300s, which can be detected from the electroencephalographic (EEG) signals and indicate which letter the user was focusing upon [22], [23]. An alternative method to spelling and writing applications leveraged motor imagery (e.g., users imagining a part of the body moving), which induces changes in EEG spectral power (often in the mu rhythm band) that are utilized to select a letter from a series of options. In such a system, an array of letters is presented to the user and the computer uses EEG signals arising from the brain's perceptual-motor system to rotate an arrow through the letter options based on which body part the user was imagining moving [24]. Despite the differences in these approaches, what they and other existing BCI approaches have in common is that they are inseparable from the task being performed (e.g. [25–29]).

The second constraint on these early applications was to target clinical populations whose inherent ability to transfer information was extremely limited, such as paralyzed or "locked in" patients. This approach proved very beneficial for enabling the direct control of items like a computer's cursor or the communication devices discussed above for these populations; however, the performance of these applications are dramatically outperformed by healthy populations using typical alternatives (i.e., a mouse for cursor control, speech for communications). In part, the reason for this is that early applications attempted to utilize the higher cortical function as a moment-to-moment control signal, thereby circumventing the highly evolved and efficient system between the brain and muscles that healthy humans normally rely upon to perform motor movements [30]. Currently, researchers are realizing the benefit of developing applications that use neural signals in ways that are more consistent with the natural neural processing for clinical populations [25], [31–33]. As researchers continue to extend BCI technologies to healthy populations, many current applications are still just extensions of the original clinical applications; however, several promising new types of applications are being developed, including attempts to integrate emotion into video games, toys, advertising, and music [34–36], as well as attempts to merge human pattern recognition with computer processing power for joint human-computer object detection [37–39].

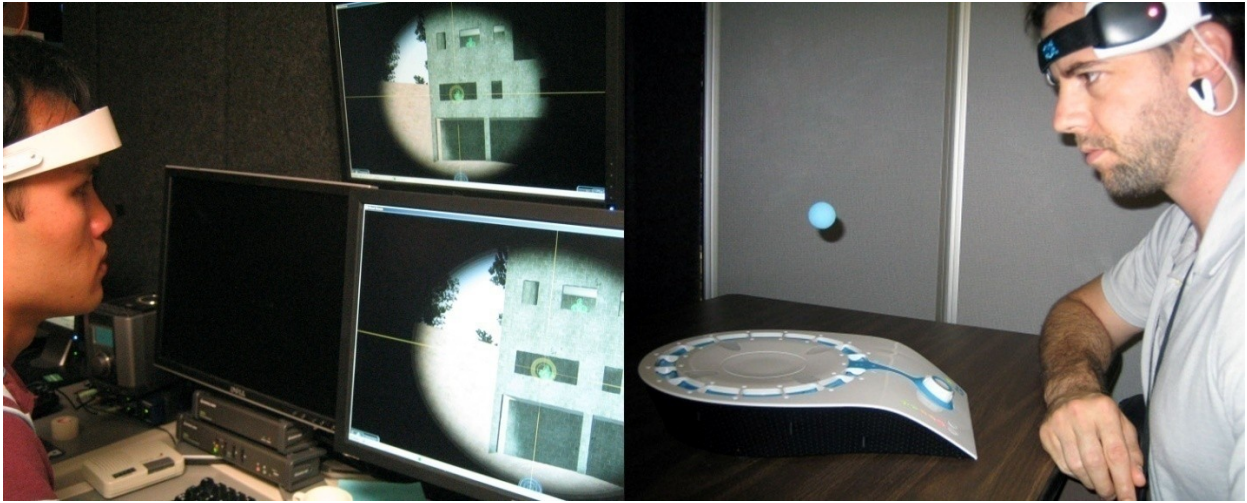

Figure 1: Brain-Computer Interfaces for Gaming. On the left a gamer uses a BCI to toggle zooming in a first person shooter game, on the right, one of the co-authors tests out his ability to use a BCI to adjust the speed of the system's fans to control the height of a ball floating on air above a platform.

*Recent Advancements in Neurotechnologies*

Over the past five years, the bridging of technological gaps in brain imaging and sensing have led to the development of the new augmented BCI (ABCI) concepts, which Liao and colleagues (this volume) define as BCIs that can be used by individuals in everyday life. As a result, ABCIs must function while people move and interact with their environment; allow non-intrusive and rapid-setup EEG solutions that require minimal training; and provide stability, robustness, comfort, and longevity for accurate long-term data collection. Technological improvements have also led to advanced algorithmic approaches to analyzing and interpreting brain data gathered under noisy, real-world environments (see Makeig et al., this volume), enabling an explosion of BCI research [40] and technology development even to the point of the commercialization of the first neurally-based toys, such as the Star Wars Force Trainer™ by Uncle Milton or the Mindflex™ by Mattel (see Figure 1). Over the next decades, neurotechnologies will increase or create a new sensing capabilities and the ability for sensors to be seamlessly integrated into user clothing and environmental devices (see Liao et al., this volume), and analytic and interpretation algorithms will be able to reliably extract user performance, self-assessment, brain states, and intent on a moment-to-moment basis (see Makeig et al., this volume), which will be enabled by the ever growing computational infrastructure. These projected developments will move brain-based neurotechnologies from toys and prototype interfaces for specialized populations to a core technology that has the potential to revolutionize human-system interactions across all aspects of everyday life.

## 3. The Promise of Incorporating the Brain

The foundation of the technological promise of BCI concepts lays in the notion that brain activity can provide unique insights into people and their behavior, and that these insights can be used to develop systems that can change how humans interact with the world. For example:

- As the nervous system underlies human behavior, the central element of the nervous system, the brain, holds vastly more information than can be deciphered through behavior alone. The wealth of additional information gained through leveraging the neural signatures provides the potential to develop fundamentally different human-computer interaction capabilities than are seen with current technologies.
- The processes of the human brain are highly variable, both across people and within an individual across time, and this variability underlies the variability observed in human behavior. As such, understanding and leveraging this neural variability may be useful for tailoring adaptive technologies to the individual user and their current mental state.
- The human brain is highly adaptable in specific ways, which enables a wide variety of human capabilities such as learning, adjusting to new tasks and environments, and even overcoming many types of trauma. Understanding how the human brain adapts and tracking neural adaptation online may be useful for leveraging this inherent human capability to develop novel approaches to training, education, and rehabilitation.

Such insights into brain processing could be merged with expected advancements in computing and artificial intelligence to move beyond the mere transmission of information between man and machine, and towards mutually-derived analysis, interpretation, and prediction of situations. That is, the combined human-computer system may be able to sense and integrate information about the operators' and the computers' past and present capabilities, states, goals, and actions, in addition to the global task and environmental constraints. Having this information could lead to deeper insights into human brain function and behavior, enabling predictions about performance outcomes, and ultimately leading to revolutionary changes in the fundamental ways humans and computers interact. Further, as the human nervous system processes and stores information in ways that are very different from the current computer systems with which they interact, from a systems perspective these insights into brain function may even impact conceptions as to when and where in the overall system design human-computer interaction should occur.  However, the realization of this potential will rely on the ability to reliably extract accurate, high-resolution indices of neural processing. Here, we briefly overview a small sampling of brain processes that the cognitive neuroscience literature has identified that may have major implications for potential BCIs. Note that while future advances towards persistent, invasive sensing technologies will surely be made (see Liao et al., this volume) and significant potential exists in leveraging these technologies (for example, cognitive neural prostheses such as [41]), the brain signals and related potential BCI applications discussed here may be realizable using non-invasive methods. In the near-term, we consider these non-invasive methods to be the most practical and user-acceptable approaches to enhance human-system interactions through BCIs.

Perhaps the most highly investigated brain functions are those associated with the processing of sensory stimuli. In the visual domain alone, brain imaging methodologies have revealed a wealth of information that can be derived from neural processes associated with stimulus presentation. For example, neuroimaging techniques can show that the brain is processing visual information [42], when that processing is taking place [43], and can also give insights into the nature of the processing: where in the visual field a particular stimulus is located [44], [45], when it was perceived [46], the frequency with which the stimuli are flashing [47], whether the stimulus was stationary or moving [48], whether an image (or mental imagery of that image) was a face (fusiform face area; [49]) or a place (parahippocampal place area; [50–52]), whether images are familiar or novel [53], and can even provide a partial decoding of a specific image [54] or video [55]. Similar modality-specific functions have been revealed for other sensory modalities.

Similarly, brain imaging reveals information across a range of motor control and higher cognitive functions that could be leveraged to support brain-based applications, such as measuring when an individual is encoding or recalling information from memory [56–58] or identifying when motor planning [59], movement initiation [60], and motor imagery [61] occur. Differences between expert and novice motor skill performers [62] can also be identified from brain data. Higher-level cognitive capabilities include the detection of deception [63], [63] and the withholding of guilty knowledge [64]; consumer preference and decision making styles [65–67]; and executive functions such as conflict monitoring, error detection and the level of conscious effort during tasks [68–70]. In addition to cognitive functions, the human brain has transient states that modulate or interact with other processes and that would allow systems to adapt to the human user's changing brain state. For example, neural measures predicting performance lapses have been uncovered based on measures linked to fatigue [71–73], emotion [74], [75], arousal [76], stress [77], engagement [78], and cognitive load [79].

Neuroimaging has extensive roots in medicine and neural indices have also been associated with various aspects of a wide range of brain injuries, disorders, and diseases. We envision that this research could support technologies that span the spectrum of medicine including: preventative or secondary preventative applications; novel diagnosis tools; treatment, restorative, or rehabilitative tools and devices, and even applications for improving quality of life and coping with ailments. Just a small sampling of the disorders and diseases for which neural indices have been associated include: epilepsy [80], [81], attention-deficit disorder [82], autism [83], mild and severe traumatic brain injury [84–86], post-traumatic stress disorder [87], [88], insomnia [89], chronic pain [90], Alzheimers [91], and addiction [92].

The information contained within the brain processes outlined above represents just a small window into the explosion of neuroscience research that has occurred over the past several decades. Currently, the abundance of developing computing systems and neurotechnologies present a remarkable opportunity to synthesize and leverage this growing knowledge base for improving human-system design. However, the development of these applications will depend on numerous factors spanning from the reliability of the neural signal sensor to the commercialization of the final product. For example, most recent neuroscience research yields correlation-based findings derived from groups of individuals and/or across numerous individual trials. While this research base illustrates the wide scope of information that can be extracted from brain signals, the use of the established techniques for extracting such information will generally be unsuccessful in the broad range of BCI applications foreseen here. However, current efforts focused on developing algorithms for real-time information extraction on individual operators (see Makeig et al this volume) and the continued development of initial BCI applications that relax the accuracy requirement of brain-derived information (e.g., lowering required accuracy rates to 80%) and/or relaxing the time constraints on the information (e.g., allowing information to be extracted from 5-10 minutes of brain signal) will help to enable the use of this broad base of neuroscience measures in BCI design. In addition, most recent neuroscience-based understandings have been derived using highly-controlled experimental scenarios that capture neither the scope nor the complexity of tasks performed in the real-world. Given the complexity and marked moment-to-moment variability of brain dynamics and the brain's central role in producing real-world behavior, the realization of the full potential of BCIs within the coming decades will be nontrivial. However, current efforts to extend neuroscience research to naturalistic tasks and real-world environments, the development of enhanced and alternative techniques for analyzing and interpreting neural signals in complex environments, and the extremely broad scientific base already established all provide the promise that the field of BCIs will continue to rapidly advance in the near future.

## 4. The Future of BCI Technologies

While there is incredible potential for the development of future BCI applications waiting to be unlocked in the hundreds of indices of neural behavior that have been identified by the neuroscience research community, current and likely near term BCIs remain "task-oriented" (i.e., where the application is directly oriented towards the task the user is trying to accomplish) and include: a) BCIs that are the primary interface for the task the user is explicitly performing, such as using brain signals to control the movement of a prosthetic; and b) BCIs that directly support the task the user is performing but are not the primary interface, such as a system that monitors the user's brain signals in order to predict performance while driving and to mitigate periods of predicted poor performance. Developers have and will likely continue to find success with task-oriented BCIs, where the application itself is controlling the conditions under which the user performs, as opposed to attempting to find brain indices that generalize across any task that a user may be performing. This is because task-oriented BCIs will have access to more context for what the user is actually doing, and thus greater capability for interpreting the incoming neural signals.

Future task-oriented BCIs, based on advances in sensor technologies, analysis algorithms, artificial intelligence, multi-aspect sensing of the brain, behavior, and environment through pervasive technologies, and computing algorithms, will be capable of collecting and analyzing brain data for extended time periods and are expected to become prevalent in many aspects of daily life. If and when brain-sensing technologies are worn during portions of people's daily lives, the possibility of using the BCI infrastructure for "opportunistic" applications arises. That is, once users are regularly wearing brain sensors for specific purposes, opportunistic BCIs, which are BCI technologies that provide the user with a benefit, but do not directly support the task the user is performing, can be employed without additional overhead. Example opportunistic BCIs could be pervasive computing applications [93], [94] that adjust the user's local environment (such as the color of lighting, music, or perhaps even odor, or suggestions for dietary, exercise, entertainment, or treatment options) to alter or enhance the user's mood or mental state, or medical applications that periodically screen the user for indicators of neural diseases and pursue a variety of mitigations. Such mitigations may include: generating tasks for further analysis and screening (moving the BCI into the task-oriented domain), suggesting the user see a doctor for diagnosis, or suggesting preventative measures. However, due to the lack of constraints under which such applications have to function, opportunistic BCI development will likely advance through large-scale collection and analysis of data over extended periods of time, as well as the development of techniques for extensive individual customization to the user. While these issues will limit near-term development, over the longer-time frame, opportunistic BCIs may have life-saving ramifications in addition to the many other potential benefits to medical, education, work, and social applications.

In the next few sections, we intertwine a discussion of several potential near- and far-term BCI applications with the factors that will have to be addressed for these technologies to achieve widespread utilization in society. Importantly, the types of applications discussed below are not independent; for example, the advancement of communications applications will likely be intertwined with and improve direct and indirect control applications.

*Direct Control*

Some of the earliest concepts for brain-computer interface applications focused on conscious direct control, i.e. using brain signals to directly manipulate the state of an object. Examples come from clinical applications (such as wheelchairs, prosthetic devices, communication applications) and from the first brain-control games (see Figure 1). In the near future, consumer demand is likely to continue to push BCIs, particularly in entertainment and quality-of-life applications, to pursue direct control. From a clinical standpoint, there are populations of patients who would greatly benefit from the ability to consciously control their own movements (e.g., wheelchairs, limbs) and devices in the world around them (lights, radios, televisions, espresso machines, computers, phones). From an entertainment standpoint, humans have a fascination with the ability to control objects by directly using their mind, demonstrated by the point that these concepts, such as telekinesis, are embedded in our popular culture. In part this fascination seems to stem from the concept that direct access to the brain will present us with vastly improved capabilities for interacting with the physical world. If such applications can be developed, they will have tremendous impacts on quality-of-life and work. However, numerous interrelated factors will impact the success of direct control applications, including:

- *What can the brain really do?* While the human brain is capable of phenomenally complex motor tasks, and can even physically adapt over time to support specific motor control tasks (e.g., the greatly elaborated motor cortex of highly experienced musicians [95]), it is also clear that our brains have limitations. Take the fictional example of a BCI that allows a person to control four mechanical arms in addition to their human limbs. Obtaining the benefits of these additional appendages will likely require multitasking, which is known to be difficult and tedious for humans [96]; and multi-limb coordination, which is also limited in humans without extensive practice [97]. As mentioned in section 2, the brain's cortex appears to provide only a portion of the direct control function, processing higher-level goals and allowing lower-level brainstem and spinal mechanisms to execute the fine control of the limbs. These examples point out that merely linking technology into the cortex (which many early BCIs have attempted) won't automatically enable limitless capabilities. Rather, technologies will likely advance though approaches that specifically coordinate with and augment the capabilities and limitations of brain function.

- *What are we extracting from the brain?* After determining appropriate neural signals, those signals must be detected and provided to the BCI. Neural signals are recorded using technologies that range from huge, expensive scanners that require the participant to lay motionless on their back, to relatively inexpensive, small form factor technologies that can be incorporated into a baseball hat [98]. None of these technologies can fully image what the brain is doing and most do not function in everyday environments. Further, analysis techniques are currently capable of obtaining relevant information from only a small portion of the neural signal, often requiring considerable computational processing to do so. Together, even the most advanced imaging and analysis techniques are providing only very specific and limited glimpses into the entirety of brain function. Further, the brain is non-stationary, adaptive (e.g., as the user ages, learns, and uses an application), and dynamic (e.g., changing along with the physiological fluctuations experienced during the day). Successful BCI development will rely on an understanding of the nature of the "glimpses" afforded by the particular imaging technologies and analysis techniques available in the coming decades and how those "glimpses" change to reflect natural brain adaptations.

- *Can we expect a more effective alternative?* Finally, in the development of any application it is important to understand alternative technologies. In the case of direct control, the primary alternative for healthy humans is using their highly evolved motor control system (i.e., using their own hands to operate devices). To date, the technologies for directly controlling devices using brain signals provide relatively low bandwidth and low signal-to-noise ratios. As a result, it is a nontrivial problem to enhance control for healthy individuals through the incorporation of brain signals into direct-control BCIs. For this to occur, the ability to analyze neural signals to add information above and beyond that more easily obtained through other channels (e.g., manual input) will need to be achieved.

Given these factors, the question remains as to what can be expected in the near-term. We expect further development of the types of direct control applications that are currently being pursued world-wide. In medical domains, we expect to see further progress on using brain signals as higher level "goals" or "intentions" to control prosthetic devices and wheelchairs [25], while alternative technologies or alternative behavioral or physiological signals (e.g., leveraging pectoral muscle or peripheral neural activity recorded via electromyography (EMG) to control prosthetic limbs for an amputee) will be used for the specifics of the control tasks [30]. The growth of applications in the entertainment industry are expected to be broader, with applications that allow for the fantasy of direct brain control, while accounting for the lack of effectiveness in the device (e.g., a virtual game could potentially limit ineffective or undesired movements generated by a BCI through limiting potential outcomes and modifying the laws of physics). As autonomous navigation and robotic coordination capabilities advance in the far term, BCIs may allow for the control of single or even multiple coordinated robotic devices. However, direct control BCI will always be in competition with alternative human-computer interface technologies.

*Indirect Control*

One of the fundamental concepts that will directly influence future BCIs is the use of brain indices that provide information that is not as readily or robustly available through other channels. One potential source for this information are the brain processes that are associated with the human perception of "errors;" these could be specific brain-produced error signals such as the Error-Related Negativity [99], or may be combinations of signals associated with errors, such as frustration/anger, attention/engagement, or comprehension. We foresee this information potentially augmenting control systems without the user having to engage directly in the control task to make corrections. For example, imagine a user observing a robotic arm reaching for a door handle. A human can perceive early in the process whether the robot's hand position is appropriate for manipulating the specific door/handle combination [100], while the algorithms that control the robotic arm could then select from multiple alternative handle manipulations, selecting a new manipulation style based on the error signals received from the human user. This example indirect control application accesses the neural correlates associated with the user's perceived "error" to influence the robotic controller's choice of manipulation strategy, but does not engage the operator directly in the control task. Generally, the success of this type of application will largely depend on the robustness, specificity, and timeliness of detecting signals that indicate user intent or approaching errors, and if a BCI can be developed to the point where the neurally-based human-system control strategy performs more effectively than alternatives in terms of overall performance and load on both the system and the user.

*Communications*

The communications domain offers potentially the largest potential area of impact for BCIs. Early communication BCIs were designed to enable a clinical population with little to no communications capability to generate text (section 2). In many ways these applications are similar to the early direct control applications for clinical populations. They serve a very specific purpose and will continue to have a great benefit to the specific clinical populations they were developed for. However, they likely will not extend effectively to healthy populations in their current form. Newer technologies, from naturalistic user interfaces to collaborative filtering approaches, are making revolutionary advances beyond merely enabling speech generation and moving towards the basis of communication: the ability to pass meaning between two or more parties. We envision future BCIs as part of a holistic system that leverages communications-specific technologies, as well as other technologies, such as pervasive sensing and computing. Specifically, BCIs could potentially support a holistic approach to communication applications through three concepts: 1) increasing the bandwidth between human and computer and the effectiveness of that bandwidth, 2) enhancing or predicting the comprehension of information in context, and 3) supporting the formation of ideas.

As human-computer interfaces expand beyond mouse and keyboard and into naturalistic user interfaces, the bandwidth between human and computer will increase. One of the breakthroughs of natural user interface concepts is the idea that in face-to-face human communication, much more information is transmitted than just the words being spoken. Other information, such as body language, facial expressions, and prosody, all add to the meaning of what is being communicated. Signals characterizing these aspects of human communication could be combined with neural signals providing additional insight into the neural state of the operator and thus additional contextual information to human-computer communication. As an example, neural signals have been used in the advertising domain for gauging and understanding the responses of consumers to different products [101]. BCIs using the real-time detection of emotions, frustration, or surprise [102] could enable training or educational applications to adapt in ways that could enhance the learning rates of the students [103]. Further, advanced BCIs that combine multiple sensor systems such as eye- and head-tracking with brain imaging can be envisioned that can estimate not only which display the user is looking at [104], but other factors that will influence the capabilities of the user such as user attention location and level [105], fatigue level [71], and arousal levels [106]. Such technologies could combine these estimates to generate probabilistic predictions as to the users processing of available information, which could then be used to alter the information displays to enhance the effectiveness of the communications bandwidth.

Taking these envisioned concepts a step forward, BCIs may predict aspects of the user's comprehension of the information. For example, there are neural signatures (e.g. N400; [107]) that indicate whether a word in a sentence is perceived by the user as semantically correct. Imagine a device that can track the incidences of semantic misunderstanding in a conversation, be it peer-to-peer, student-to-teacher, or even human-to-computer. With this type of information, systems are foreseen that can provide general indicators of comprehension or communication efficiency between parties. Further, systems could cue users to repeat or rephrase conversations, or even suggest alternatives to wording that was ambiguous, misleading, or incorrect. For example, social BCIs that combine such comprehension analysis systems with emotional cues from an audience could aid in the crafting of public speeches, advertisements, and entertainment.

Technologies are also developing that could enable computers to analyze and predict what users are attempting to communicate. These range from computer search engine technologies that use collaborative filtering to suggest search terms [13], to computer vision algorithms that use graph theory to find objects that are similar to a predefined set [108]. Prototype BCIs are currently being used to augment the computer vision algorithms to help operators find objects within an environment [37]. Extensions of BCIs that leverage the types of signals described throughout this section are envisioned that would aid in the formation of joint human-computer semantic lexicons (including words, images, sounds, etc.) that would be tailorable to individual users and that could form the basis of systems that give users the perception of talking to computers or other entities such as semi-autonomous robots. As is already being seen in some search engine technologies, such lexicons would provide capabilities that go well beyond just interfacing with the computer, but would enable computers to make connections between a user's concepts, and potentially provide novel ideas back to the user based on the their inputs. A step further, imagine an overseer computer that analyzed the ideas or concepts multiple people were communicating and suggested communications that would facilitate conversation for the entire group. Such systems could become a fundamental component of human-system communication and human-human comprehension.

*Brain-Process Modification*

BCI technologies can also provide the potential for users to actively modify their own brain processes or states. Methods such as neurofeedback can already allow for individuals to adjust their own brain function in an attempt to attain a more desirable state [109]. As sensor technologies and analytical approaches improve, so do the potential benefits of neurofeedback. There are multiple potential applications of these methods, although the most promising would be for training and rehabilitation. Existing research shows that it is possible to discriminate between the brain processes of novices and those of experts at various tasks that are either physically demanding or that require considerable concentration [110], [111]. Such information could be used to gauge an individual's learning status, but could also be used to train novices how to have brain processing similar to that of an expert performer. A similar concept for applying brain-process modification is the rehabilitation of neural ailments or damage, such as that caused by a stroke. Some mental ailments, particularly affective disorders such as depression [112], may be treatable through an individual modulating their mental states with help of an advanced neurofeedback-based BCI [113–115]. Finally, it may be possible to delay age- or ailment-related neural degradation through entering specific brain states [116].

In addition to neurofeedback-based approaches, future brain-process modification BCI technologies could be developed that are based on neural stimulation. While much of the current neural stimulation research is based on invasive probes [117], there are potentially valuable noninvasive neural stimulation techniques based on direct current, magnetic fields [118], ultrasound [119], or infrared light [120]. By combining future high-resolution neural stimulation technologies with an advanced neurofeedback-based BCI, neural stimulation could assist the user in achieving desired brain processing through the stimulation or suppression of activity in brain regions of interest. This potentially could drastically improve the performance of the neurofeedback-based brain-process modification. A considerably further-term (and higher-risk) possibility would be for the system to use neural stimulation to place the user into the appropriate brain state for the relevant task the user is performing, although this would require considerable improvements in neural state detection and autonomy, as well as a strong consideration of the ethics involved.

While the ability to modify brain-processes shows promise, the incredible complexity of the brain and the large between-persons variability in neural processing suggests that it may be difficult to determine optimal neural goal-states for a specific individual. Further, as neural feedback is still largely unexplored, the potential impact of this type of training on other tasks is not well understood. These factors suggest that while brain-processing modification BCIs may enhance training and rehabilitation, it is unlikely that these technologies will revolutionize these domains in the near-term.

*Mental State Detection*

One of the recent common themes in BCIs is the detection and use of mental states, as opposed to specific instances of neural processing such as an event-related potential, to modify a system [121]. As alluded to in the above sections, the ability to reliably and accurately detect fatigue, attentional, arousal, and affective levels could allow systems or environments to adapt to the state of the user, increasing joint user-system performance across a wide range of tasks [122] or helping the user achieve a desired mental or emotional state. As one example, there is a considerable amount of current research on predicting fatigue-based performance decrements while driving [71], [123]. By integrating such predictors with fatigue-mitigation techniques, life-saving BCI technologies are envisioned that can decrease the odds of catastrophic driver errors. Similar fatigue- or attention-based systems may generalize to a wide variety of similar vigilance-based tasks, where attention must be maintained over long periods of time.

Training systems offer another opportunity for mental state monitoring based BCIs. Training systems, such as intelligent or adaptive tutoring systems, are intended to act as a replacement for a human tutor, providing personalized training and feedback to develop individual proficiency at a specific skill or task [124]. These intelligent tutoring technologies could greatly benefit from state detection systems that identify brain states indicative of a lack of learning, such as fatigue, frustration, confusion, strong negative affect, or low arousal; and brain states that are associated with learning or the sufficiency of learning. In the near-term, these detection capabilities could conceivably be used in combination with behavioral and task performance data to isolate potential influences on a student's progress, and in the long run could be combined with pedagogical theories and user modeling to allow adaptation of the training system to the user's mental state in real time, improving the learning outcomes of the tutoring system.

State-based detection is also beneficial to medical applications. Diagnosis BCIs are envisioned that could reduce diagnosis time and expense. For example, in hospital fMRI-based BCIs could analyze neural-states (or even specific instances of neural processing) during task performance and in real-time alter tasking as specific diagnoses are eliminated and others are narrowed in on. Similarly, mobile EEG-based applications could allow patients to periodically perform this type of iterative examination in their own homes, which would be beneficial to a wide range of patient populations and may provide greater opportunities for cost-effective preventative medical approaches. State-based approaches could also be combined with the aforementioned rehabilitation BCIs to enhance performance.

### *Opportunistic State-Based Detection*

Once technology reaches the point where useful opportunistic BCIs can be realized; a wide range of state-based applications can be envisioned (see Figure 2). Neural state monitoring could be used in combination with pervasive intelligence to opportunistically change the environment. For example, affective state could have numerous home, entertainment, and medical applications, including providing entertainment, exercise, or food suggestions, or directly adjusting music selections. Similarly, with a physician's recommendations, early markers of the onset of ailments such as migraines could be used to trigger recommendations for medication, stopping particular activities, or could directly adjust room lighting or other environmental factors to help alleviate an individual's symptoms.

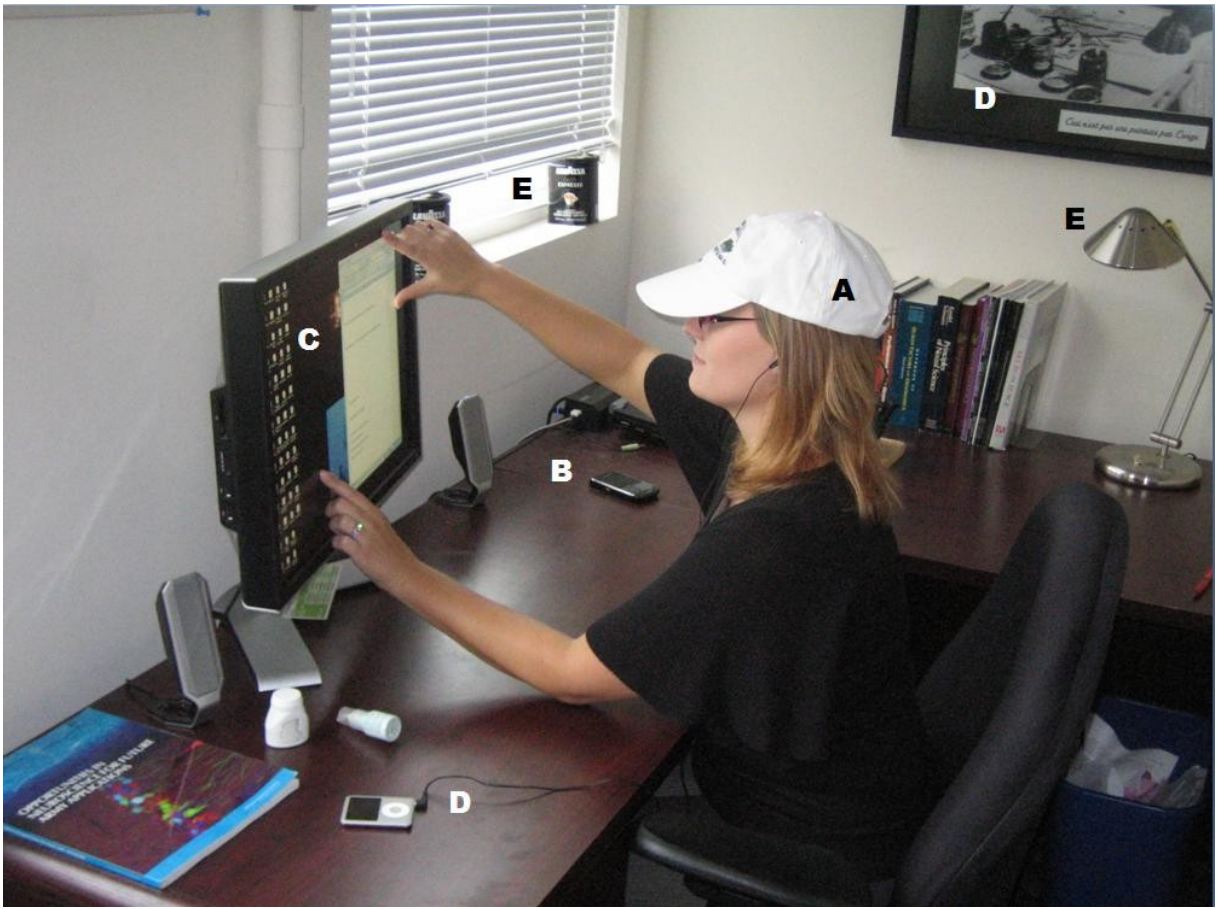

Figure 2: BCIs based on pervasive technologies embedded in an office environment. A task-oriented BCI includes: (A) dry, wireless EEG sensors embedded into a baseball cap; (B) EEG signal collection and processing on portable smartphone, which also integrates with other devices in the environment; and is used to (C) improve human-computer interaction and work performance from online detection of cognitive and affective brain states, potentially including error-related signals, emotion, and time-on-task fatigue. The use of the task-oriented BCI avails opportunistic BCIs that (D) modify music selection and image in digital frame based on online detection of affective state; (E) alter lighting and office shades due to detection of oncoming migraine through EEG, and/or C) cue the user to potential non-work activities based on lack of sleep-based fatigue or stress. In addition, longer term data collection and analysis could result in diagnostics and recommendations for neural ailments.

Fatigue and sleep-based BCIs offer another area for opportunistic applications. Most alarm clocks indiscriminately go off regardless of what stage of sleep the user is in, despite the fact that a person's energy level upon awakening is tightly linked to the sleep stage they were in just before waking [125]. By incorporating brain activity into an interactive system that wakes the user up during an optimal phase of the sleep cycle, the user could awaken feeling more refreshed and alert in the morning. Similarly, a sleep-related BCI could monitor the amount of rapid eye movement sleep, which may serve a critical function related to memory consolidation [126], to enhance memory performance. Systems like these could also monitor sleep patterns over time, or even incorporate day-to-day information from a digital calendar to suggest ideal times to sleep or the type of alarm to use on a given night.

Medical monitoring could also benefit from utilizing opportunistic brain state detection technologies (Figure 3). In the longer term, instead of providing the user with periodic tasks or exams used for medical diagnosis, the testing routines could be opportunistically derived from the individual's daily living, potentially allowing for minimally invasive testing and the earliest detection of slow onset neural pathologies, such as Alzheimer's disease. Such technologies also could make possible increased frequency of brain monitoring for rehabilitation patients or to support at-home care, potentially making higher quality medical care and independent living easier for clinical or elderly populations. For example, pervasive brain monitoring applications that could detect the onset of clinically-relevant symptoms could be coupled with automated, remote, active treatment modalities to minimize or even prevent the onset of harmful or even deadly conditions, such as epileptic seizures. Moreover, these types of applications could mesh well with technologies like virtual medical agents (e.g. [127]), in particular for applications such as stroke rehabilitation. An agent like this could utilize both task-oriented and opportunistic brain state monitoring systems to provide the patient with periodic neuropsychological rehabilitation treatments or evaluation tasks, while also monitoring the patient's progress during the performance of real-world tasks as part of their day-to-day living.

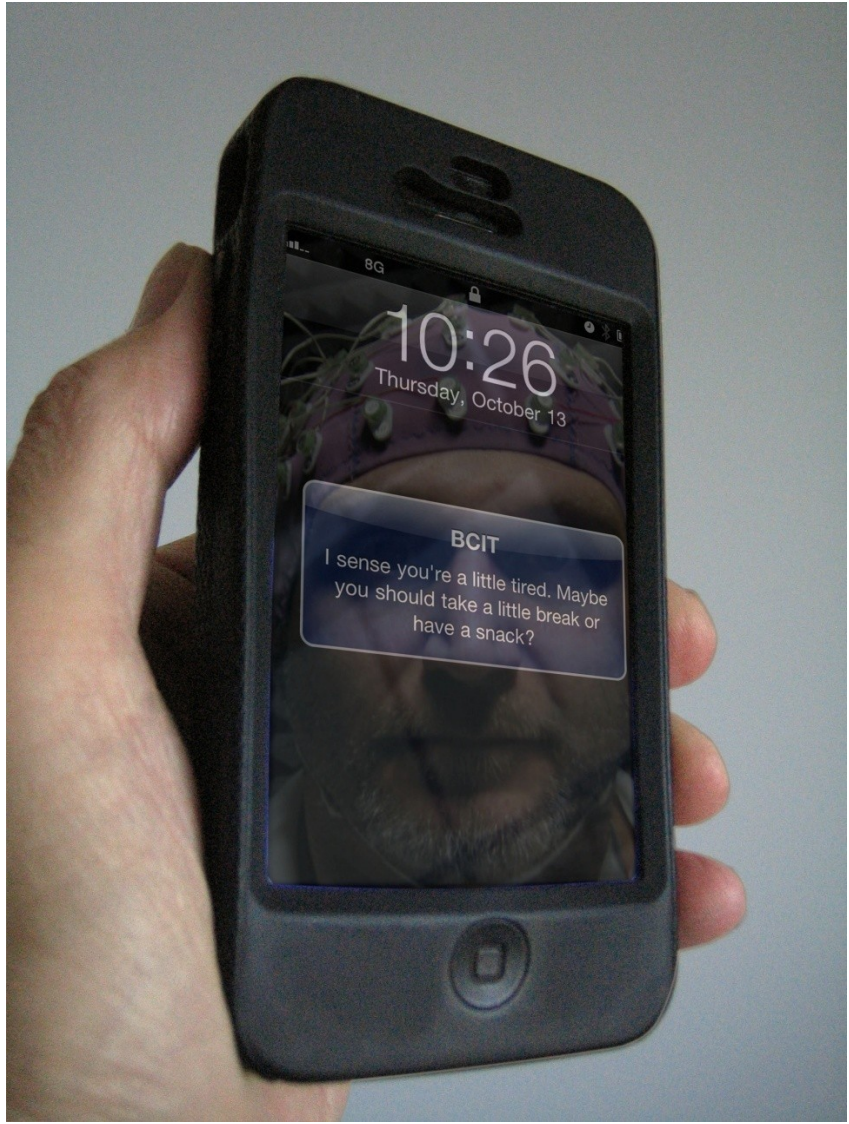

Figure 3: An example of a potential opportunistic medical diagnosis-based BCI. Based on diagnosis by a doctor (e.g., depression), a BCI could be used to detect particular neural states or activities and politely suggest activities to alter the mental state or lifestyle of the patient.

While neural state detection applications face many of the same developmental issues faced by other applications, the benefits of such technologies - in task-oriented and opportunistic applications – may provide ways to both improve human-system task performance, and to develop assistive medical applications for individuals with neural ailments, with the potential to provide numerous benefits to both healthy and clinical populations. In addition, by integrating these technologies with pervasive computing or assistive agent technologies, it may become possible to achieve vastly improved outcomes.

5. **Conclusions**

The current explosion of neuroscience research and neurotechnologies provides the opportunity to provide computers predictive capabilities for the emotional and cognitive states and processes of the people using them, potentially revolutionizing not only interfaces, but the basic interactions people have with these systems. However, to reach their full potential, the development of BCI technologies over the coming decades will have to overcome a number of obstacles. For example, the amazing abilities of people to adapt to dynamic, complex tasks and environments present difficulties in interpreting an individual's neural processes and behavior at any given time. These difficulties may arise due to the signal noise caused by environmental effects, overlapping neural processes arising from the performance of multiple concurrent tasks, and changes in neural signatures over the short and long term, in addition to the wide variation in neural signals across individuals.

In order to address these obstacles, near-term applications are likely to be task-oriented, focusing on applications where neural signals can provide information that is difficult or impossible to obtain through other measures, where perfect performance is not required for the application to successfully produce value, and that emphasize application-specific performance instead of attempting to detect abstract constructs (i.e., attempting to predict performance declines at specific tasks over time, instead of attempting to predict general fatigue). Near-term applications are also more likely to be successful if they focus on the individual user, through calibration or individual-based classification algorithms, instead of attempting to perform across broad groups or utilize normative populations [128], [129].

In the far-term, we envision a more holistic approach to BCIs that merges critical brain, behavioral, task, and environmental information obtained with advanced pervasive, multi-aspect sensing technologies, sophisticated analytical approaches, and enabled by advances in computational infrastructure such as extensions of cloud technologies. Such an approach may also benefit from exploring synergies between the human and the computer as well as the large-scale collection of data consisting of both brain function (e.g. EEG, fMRI) and brain structure (e.g. diffusion weighted imaging [130]) at multiple scales, ranging from individual neurons up to maps of the entire brain. This data could provide a great deal of insight into how differences and changes in physical brain structure, both within and between individuals, cause changes in the functional brain data that can be detected in real time, thus providing much greater capabilities to individualized BCI technologies. The pervasive integration of neurotechnologies will also avail the development of a broad range of opportunistic BCI technologies in the far term, which have the potential to dramatically influence quality of life on a daily basis if scientists and developers can overcome the hurdles associated with detecting and interpreting neural signatures in relatively unconstrained settings.

In this paper, a number of potential BCI technologies focused on communication and other applications have been described; however, these represent just a small sample of the broad future potential of these technologies. We have also focused the discussion of applications on relatively foreseeable breakthroughs in sensor, analysis, and computational technologies; however, unforeseen breakthroughs, such as a novel wearable sensing technology that provides ultra-high resolution, real-time imaging of both the spatial and temporal activities of the brain, would open the door to vastly wider set of applications.

Acknowledgements:

We would like to thank the four reviewers, as well as Christian Kothe, Tzyy-Ping Jung, and Chin-Teng Lin for their insightful comments on the paper. We would also like to thank Paul Sajda for our valuable discussions on future neurotechnologies.